\documentclass{article}
\usepackage{spconf,amsmath,graphicx}
\usepackage{amsfonts,amssymb,amsthm}
\usepackage{algorithm}
\usepackage{algorithmic}
\usepackage[numbers]{natbib}
\usepackage[graphicx]{realboxes}
\usepackage{verbatim}
\usepackage{algorithm}
\usepackage{algorithmic}
\usepackage{amsmath}
\usepackage[hidelinks,colorlinks,linkcolor=blue,filecolor=blue,citecolor=blue,urlcolor=blue]{hyperref}
\usepackage{xcolor,soul}
\usepackage{tabularray}
\usepackage{listings}
\usepackage{bm, diagbox, booktabs}
\usepackage{subcaption}
\usepackage{url}
\usepackage{pdfpages}

\title{Extreme Video Compression With Prediction \\Using Pre-trained Diffusion Models}
\name{Bohan Li$^{\star}$, Yiming Liu$^{\star}$,  Xueyan Niu$^{\ddagger}$, Bo Bai$^{\ddagger}$, Lei Deng$^{\ddagger}$, and Deniz Gündüz$^{\dagger\ddagger}$}
\address{$^{\star}$Xidian University, Shaanxi, China, \{bohanli, yimingliu\}@stu.xidian.edu.cn\\
$^{\dagger}$Imperial College London, London, U.K, d.gunduz@imperial.ac.uk\\
$^{\ddagger}$Huawei Technologies Co. Ltd., \{niuxueyan3, baibo8, deng.lei2\}@huawei.com}

\begin{document}
%
\maketitle

\begin{abstract}
  Diffusion models have achieved remarkable success in generating high quality image and video data. More recently, they have also been used for image compression with high perceptual quality. 
    In this paper, we present a novel approach to extreme video compression leveraging the predictive power of diffusion-based generative models at the decoder. The conditional diffusion model takes several neural compressed frames and generates subsequent frames. When the reconstruction quality drops below the desired level, new frames are encoded to restart prediction. The entire video is sequentially encoded to achieve a visually pleasing reconstruction, considering perceptual quality metrics such as the learned perceptual image patch similarity (LPIPS) and the Fréchet video distance (FVD), at bit rates as low as 0.02  bits per pixel (bpp). 
    Experimental results demonstrate the effectiveness of the proposed scheme compared to standard codecs such as H.264 and H.265 in the low bpp regime. The results showcase the potential of exploiting the temporal relations in video data using generative models.
    Code is available at: \href{Extreme-Video-Compression-With-Prediction-Using-Pre-trainded-Diffusion-Models-}{https://github.com/ElesionKyrie/Extreme-Video-Compression-With-Prediction-Using-Pre-trainded-Diffusion-Models-}

\end{abstract}
\section{Introduction}

\label{sec:intro}
Recent years have witnessed an exponential growth in demand for video data, which contain richer spatial and temporal information than other information sources such as image, audio and text. With the emergence of augmented and virtual reality (AR/VR) technologies and metaverse applications, the need for highly compressed video transmission is expected to become ever more crucial. Over the past decades, many video compression methods have been standardized and commercialized, notably H.264/AVC, 
H.265/HEVC,
and H.266/VVC. 
These compression methods are often hand-engineered with techniques such as intra-frame and inter-frame compression and entropy coding. 
The rapid advances in artificial intelligence in recent decades have brought new opportunities to video compression, and many methods have been shown to perform on par or better than traditional codecs by replacing certain components with learned models \cite{rippel:ICCV:19, Veerabadran:CVPRW:20, Jin:TIP:23}. Both the traditional methods and the neural compression techniques rely on optical flow and motion vector estimation to capture motion information, which is then used for frame interpolation, followed by residual compression and entropy coding.

In parallel, there have been significant progress in generative AI technologies, and their applications to video generation, prediction and in-filling \cite{Blattmann:CVPR:23, ho2022video, luc:CoRR:20, Xu:WACV:20}. 
In this paper, we propose a video compression scheme relying on the generative power of diffusion models \cite{Ho:NeurIPS:20, song2021scorebased}. The core idea of the diffusion model is to simulate the information propagation process among pixels in an image by gradually adding Gaussian noise. The network captures the low-dimensional representations, and the reverse process uses the learned features to generate diverse high-quality reconstructions. We use the diffusion model to decode certain video frames by recursively predicting future frames conditioning on encoded frames within specific time windows. The compression scheme ensures that the generated video frames achieve a pre-defined reconstruction quality. A subset of chosen frames is chosen based on the required quality and encoded using neural image codecs. Then they are passed as conditional frames to the video generation model, which generates several subsequent frames. Since only a subset of original frames needs to be encoded and the remaining frames can be generated by prediction, the scheme can achieve an ultra-low bit rate (below 0.02 bpp) with comparative visual quality relative to other standard video compression methods that require optical flow estimation and motion compensation.

\section{Related Work}
\label{sec:relatedworks}
    \subsection{Video Compression Codecs}
    Traditional video compression techniques, such as AVC, HEVC, and VVC, are mostly hand-engineered. In recent years, learning-based video compression has emerged as a new research direction. 
    The attention mechanism used in motion compensation by the deep contextual video compression (DCVC) scheme in \cite{li2021deep}  is proved to be more effective than the optical flow reverse warping technique employed by the DVC scheme proposed in \cite{lu2019dvc}. 
    The authors of DCVC have further improved their work and ultimately proposed DCVC-DC \cite{li2023neural}. 
    We will use DCVC-DC as one of the baselines for comparison with the performance of our model.
    These works have demonstrated remarkable performance, outperforming traditional compression methods like H.264 and H.265 in terms of PSNR and MS-SSIM metrics. 
    
    
\subsection{Video Prediction and Generation}
Video prediction and frame interpolation aim to generate new video frames according to existing ones preserving the spatial and temporal coherence. Traditionally, methods such as optical flow estimation and motion vector estimation have been developed to capture motion information which is then used for frame interpolation. Data-driven approaches using GANs and diffusion models have shown promising performance gains in real-world data synthesis.
GANs for video generation and prediction have gone through a remarkable development over the last few years and produce impressive results \cite{tulyakov2018mocogan}. 
%
Diffusion models outperform GANs in an increasing number of tasks and diverse domains and have also been deployed in video generation \cite{kawar2022denoising,yang2022diffusion,hoppe2022diffusion}. Nevertheless, diffusion models are in general computationally more intensive, and as the prediction step increases, the quality of the synthesis deteriorates. In our work, we address this issue by using an autoregressive approach, drawing inspiration from a diffusion-based model called masked conditional video diffusion (MCVD) \cite{voleti2022mcvd}, which processes relatively short video segments each time and iteratively produces the frame generations. MCVD employs a score-based diffusion model, and has been shown to achieve state-of-the-art (SOTA) results on various datasets including SMMNIST, KTH, BAIR, and Cityscapes, while being computationally efficient. 
For the denoising network, an enhanced U-Net architecture is utilized, integrating 2D convolutions, multi-head self-attention mechanisms, and adaptive instance normalization. MCVD applies a conditioning procedure based on masking past and/or future frames in a blockwise manner. 
The model generates videos in fixed-step increments, using the generated video as conditioning frames for subsequent generations. 
MCVD also supports a variety of video tasks, including future/past prediction, unconditional generation, and interpolation.

\subsection{Generative models for video compression} 
Generative models have been employed for video compression to obtain better perceptual quality at the decoder. Mentzer et al. \cite{Mentzer:ECCV:22} employ GANs in conjunction with the neural compression pipeline used in \cite{lu2019dvc} for improved perceptual quality.
While these works provide certain improvements, they do not exhibit gains in terms of LPIPS or FID compared to H.264/H.265.
To the best of our knowledge, there is no prior video compression framework that benefits from the power of diffusion models, despite highly promising results for image compression \cite{yang2023lossy, ghouse2023residual, Hoogeboom2023a}.

\begin{figure}[t]
\centering
\includegraphics[width=0.5\textwidth]{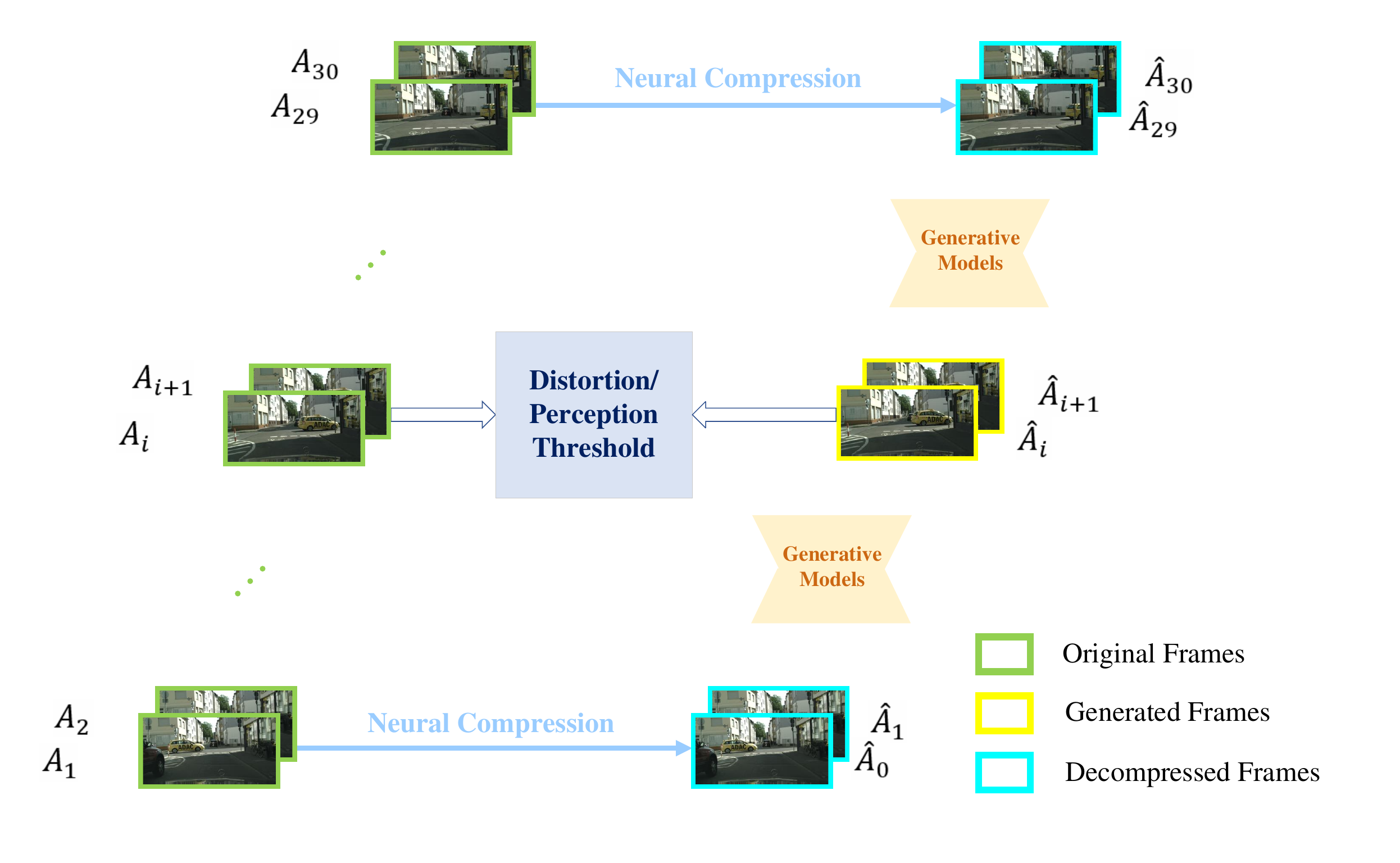}
\caption{\textbf{Method overview.} 
The first few frames are compressed by the encoder, while the following frames are generated using a pre-trained generative model at the decoder.  When the generation quality drops below the desired threshold, new frames are encoded to sustain the overall visual quality.}
\label{fig:framewrok}
\end{figure}

\section{Methods}
\label{sec:methods}

The proposed predictive video compression framework consists of two components functioning in an iterative manner: (i) compression of individual frames using neural image compression; and (ii) frame generation using diffusion-based generative models. 
Thanks to the power of the pre-trained generative model, only a subset of the frames are compressed using neural image compression, while the remaining frames are generated by the pre-trained generative network at the decoder, and hence, require no communications. 
See Fig.~\ref{fig:framewrok} for a high-level depiction of our method.

\subsection{Preprocessing}
Let $A_{1:T}\in \mathbb{R}^{T\times C\times H\times W}$ be a sequence of video frames, where $T$ denotes the number of frames, $C$ denotes the number of channels ($C=3$ for RGB), and $H, W$ denote the height and width of the frames. We assume that $A_t$ evolves over time following the joint distribution $p(A_{1:T}).$

In the proposed framework, the initial $k$ frames $A_{1:k}$ and a subset of intermediate frames are encoded using a SOTA image compression method. In our implementation, we use the SOTA neural compression scheme ELIC \cite{he2022elic} for compression of individual frames.


\subsection{Frame generation with diffusion-based models}

In standard video coding, and all the neural-based video compression schemes, frames between intra-coded frames are coded in reference to the intra-frames, typically through motion or optical flow estimation. Instead, in the proposed approach, we rely completely on the generative power of the diffusion-based neural networks at the decoder. Let $S\subset [T]$ be the subset of intra-coded frames compressed individually. Then, the remaining frames, for example $A_i\in A_{[T]\setminus S},$ will be generated using a pre-trained diffusion model with parameter $\theta$, conditioned on the previous $k+1$ frames $A_{i-k-1:i-1}.$ 

We assume that a model $p(A_{t+k+1}|A_{t:t+k})$ is learned by the generative model during the training process. Specifically, let $A_0 \in \mathbb{R}^d $  be a sample from the data distribution $p_{data}$. The forward diffusion process (FDP) with defined variance schedule $\beta_t$ from $t=0$ to $t=T$ is obtained by adding Gaussian noise terms
\begin{equation*}
q(A_{t}|A_{t-1}) =\mathcal{N}(A_t;\sqrt{1 - {\beta_t}} A_{t-1}, \beta_t \mathbf{I}).
\end{equation*}
The reverse diffusion process (RDP) is also defined as a Markov chain with learned Gaussian transitions starting from  $A_T \sim \mathcal{N} (0,I)$, which is computed using the transition kernel
\begin{equation*}
 p_\theta (A_{T\colon0} | y)= p(A_T) \prod_{t=1}^T p_\theta (A_{t-1} | A_t,A_0)  
\end{equation*}
 with $p_\theta (A_{t-1} |A_0) = \mathcal{N} (A_{t-1};\mu(A_t,A_0),\widetilde{\beta_t}\mathbf{I})$, where $\bar{\alpha_t}\colon=\prod_{s=1}^t (1 - \beta_s)$,  
 $\widetilde{\beta}_t = \frac{1-\bar{\alpha}_{t-1}}{1-\bar{\alpha}_{t}}\beta_t$, and $\mu_t(A_t, A_0)=\frac{\sqrt{\bar{\alpha}_{t-1}}\beta_t}{1-\bar{\alpha}_{t}}A_0 + \frac{\sqrt{\bar{\alpha}}_{t}(1-\bar{\alpha}_{t-1})}{1-\bar{\alpha}_{t}}A_t$.

Given a sequence of past $k+1$ frames $A_{i:i+k}$, a video prediction model, for example, as proposed in \cite{voleti2022mcvd}, is trained to learn the conditional distribution of $p(A_{i+k+1}|A_{i:i+k})$ with the objective
$L_{\text{pred}}(\theta) = $
\[
\mathbb{E}_{t, [\mathbf{y}, \mathbf{x}], \boldsymbol{\epsilon} \sim \mathcal{N}(\mathbf{0}, \mathbf{I})} [ \Vert \boldsymbol{\epsilon} -
\boldsymbol{\epsilon}_{\theta}\left(\sqrt{\bar{\alpha}_{t}} \mathbf{x} + \sqrt{1-\bar{\alpha}_{t}} \boldsymbol{\epsilon} \mid \mathbf{y}, t\right) \Vert_{2}^{2} ].
\]
\subsection{The sequential encoding process}

The key mechanism of the proposed video compression framework is a decision process at the encoder that strategically excludes a substantial portion of frames according to the pre-trained generative model without compromising the quality of the reconstructed video frames. 
The pseudo-code of the procedure is presented in Algorithm~\ref{pseudocode}.

During the encoding, a window of $k+1$ frames $A_{i:i+k}$ is processed at once, with $i$ ranging from 1 to $T-k.$ 
A list $S\subseteq [T]$ of frame number is maintained to track if the current frame is encoded or if it will be generated at the decoder using the previous frames. 
We initialize this list with the first $k$ frames, such that $[k]\subseteq S.$
Specifically, at time $t>k,$ an estimation $\widetilde{A}_{t+1:t+j}$ is derived using the given generative model, which takes as input the frames $\widetilde{A}_{t-k:t}$ and predicts the next $j$ frames 
\begin{equation*}
\begin{split}
\widetilde{A}_{t+1:t+j} &= \arg\max p_{\theta}(A_{t+1:t+j}|\widetilde{A}_{t-k:t})\\
                        &=:G_\theta (\widetilde{A}_{t-k:t})
\end{split}
\end{equation*}
according to the learned conditional distribution $p_\theta$. The result is then compared to the uncompressed data with a given distortion/perception threshold $\rho>0$. Let $\mathcal{D}(\cdot, \cdot)$ be the similarity metric. 
For each frame of the generated frames, if it meets the thresholding quality requirements, i.e., if $\mathcal{D}(A_{t+i}, \widetilde{A}_{t+i}) < \rho$, where $0 \leq i \leq j$, we remove the frame by adding $t+1$ to the set $[T]\setminus S.$ Otherwise, if $\mathcal{D}(A_{t+i}, \widetilde{A}_{t+i}) \geq \rho$, where $0 \leq i \leq j$, we add $[t+1:t+k]$ to the set $S$, which means that the frame $A_{t+1:t+k}$ is encoded using the codec. The procedure is conducted sequentially until $t=T-1$.
During decoding, the frames $A_{S}$ are decoded using the corresponding decoder of the codec used for compressing intra frames. For the remaining frames, each $\hat{A}_i, i\in [T]\setminus S$ is generated by the pre-trained generative network taking $\hat{A}_{i-k-1:i-1}$ as condition. The thresholding process ensures that the quality of the aggregated decompression meets the given quality requirement $\rho$ concerning the metric $\mathcal{D(\cdot,\cdot)}.$

\begin{algorithm}[tbh]
    \caption{Sequential encoding with codec and pre-trained neural network}\label{pseudocode}
    \textbf{Input}: Original video frames $A_{1:T},$ Quality threshold $\rho$, \\
    Neural Image encoding method $\mathrm{Enc}(\cdot)$, Neural Image decoding method $\mathrm{Dec}(\cdot),$ \\
    Pre-trained generative model $G_\theta(\cdot)$\\
    \textbf{Parameters}: Length of video $T$, Lengths of generation windows $k$ and $j$\\
\textbf{Output}: List of frame index $S$  , List of frames $\mathcal{A}$ encoded using neural codec

    \begin{algorithmic}[1]
        \STATE $S \gets [k]$, $\mathcal{A} \gets \mathrm{Enc}(A_{1:k})$, $l \gets k$, $m \gets k$
		\WHILE{$l \leq T-1$}
            \STATE $ \widetilde{A}_{l+1:l+j}= G_\theta(\widetilde{A}_{l-k:l})$
            \FOR{$ i = 1$ {\bfseries to} $j$}
    		\IF{$\mathcal{D}(\widetilde{A}_{l+i}, {A_{l+i}}) < \rho$ } 
                    \STATE $\mathcal{A} \gets \mathcal{A} \cup \{*\}$
    		      \STATE $m\gets m+1$ 
    		\ELSE
    		      \STATE $S \gets S \cup [l+i:l+i+k]$
    		      \STATE $\mathcal{A} \gets \mathcal{A} \cup \{\mathrm{Enc}(A_{l+i:l+i+k})\}$
                    \STATE $m\gets m+k$
            
                    \STATE $Break;$ 
    		\ENDIF 
            \ENDFOR
            \STATE $l \gets m$
		\ENDWHILE 
    \STATE \textbf{return} $S, \mathcal{A}$
    \end{algorithmic}
\end{algorithm}

\section{Experimental Results}

\subsection{Experimental Setup}

\textbf{Dataset}: We conducted experiments on three datasets: Stochastic Moving MNIST (SMMNIST) dataset \cite{srivastava2015unsupervised}, Citys-capes dataset \cite{Cordts2016Cityscapes}, and Ultra Video Group (UVG) dataset \cite{mercat2020uvg}. The first two datasets are widely used in the domain of video prediction, while the UVG dataset has found extensive applications in the field of video compression.
The SMMNIST dataset is an extension of the MNIST dataset, where the digit images form a temporal sequence, creating dynamic scenes as a result of random movements of the black-and-white digits, and we use the original resolution of $64 \times 64$. 
The cityscapes dataset contains high-resolution images of different city streets recorded under various conditions, simulating the diversity of real world. The dataset has already been split into training and testing, so we randomly sample 720 frames from the test set. 
We  center-crop and down-sample the original frames to a resolution of 128$\times$128. In our experiments, each video consists of 30 frames. The UVG dataset comprises 16 video sequences of 3840$\times$2160. 
For our experiments, we selected 7 videos with a frame rate of 120 frames per second (fps). Similar to the Cityscapes dataset, we resized these video frames to 128x128 and took the initial 30 frames.

\textbf{Evaluation Metrics}: Conventional metrics for video compression, such as the peak signal-to-noise ratio (PSNR) and the structural similarity index measure (SSIM), evaluate the distortion of the reconstructed video frames at the pixel level. In recent years, an increasing body of research has demonstrated that these quantitative metrics are insufficient in capturing the reconstruction quality perceived by humans \cite{blau2019rethinking}. In our study, besides PSNR, we employ the learned perceptual image patch similarity (LPIPS) metric \cite{Zhang_2018_CVPR} as well as the Fréchet video distance (FVD) metric \cite{unterthiner2019fvd} that have been popular for evaluating the perceptual qualities of generated video sequences. 
For all the videos, we computed the average metrics at the various bpp levels and provided a rate-distortion (perception) assessment of the compression performance.
We report the average values and variances of the evaluation metrics.

\begin{table}[t]
    \centering
    \resizebox{0.5\textwidth}{14mm}{
    \begin{tabular}{c|c|c|c}
    \toprule 
    Methods & PSNR  ($\uparrow$)       & LPIPS ($\downarrow$)          & FVD ($\downarrow$)            \\
    \midrule 
    H.265                          & $22.44 \pm 2.18$ & $0.22 \pm 0.04$&  $3886.70\pm 1174.39 $  \\
    H.264                          &  $24.74\pm2.48$  & $0.13 \pm 0.04$   &  $2414.48 \pm 830.35$\\
    Ours                   &  $24.37 \pm 2.57 $  &  $0.10\pm 0.03$   &  $\bm{737.96}\pm 274.48 $  \\
    DCVC-DC &  $\bm {34.68 }\pm 2.01 $  &  $\bm{0.04}\pm0.02$   &  $745.86\pm 401.16 $  \\
    \bottomrule 
    \end{tabular}
    }
    \caption{Comparison of PSNR, LPIPS, and FVD for ground-truth videos and different approaches. We report metrics at $\mathrm{bpp}=0.06$ here. Note that H.265 cannot reach bpp as low as 0.05. Similarly, DCVC-DC also fails to compress to a bitrate of 0.05 in certain videos.}\label{table}
\end{table}

\begin{table*}[t]
    \vspace{0.3in}
    
\resizebox{\textwidth}{!}{%
\begin{tabular}{|c|c|c|c|c|c|c|c|c|c|c|c|c|c|}
\hline
\diagbox{Data}{Metrics}  & \multicolumn{4}{c|}{PSNR} & \multicolumn{4}{c|}{LPIPS} & \multicolumn{4}{c|}{FVD} \\
\hline
Bpp=0.06 &Ours & H.264 & H.265 & DCVC-DC & Ours & H.264 & H.265 & DCVC-DC & Ours & H.264 & H.265 & DCVC-DC\\
\hline
UVG-YachtRide & 25.61 & 26.44 & 23.95 & \textbf{35.20} & 0.096 & 0.075 & 0.14 & \textbf{0.009} & 2540 & 1958 & 4282 & \textbf{218} \\
UVG-Beauty & 28.45 & \textbf{28.76} & 25.28 & nan & \textbf{0.057} & 0.086 & 0.17 & nan & \textbf{1416} & 1913 & 3227 & nan \\
UVG-Bosphorus & 24.55 & \textbf{29.57} & 26.70 & nan & 0.101 & \textbf{0.053} & 0.104 & nan & \textbf{2079} & 2275 & 2951 & nan \\
UVG-HoneyBee & nan & \textbf{25.37} & 21.55 & nan & nan & \textbf{0.047} & 0.196 & nan & nan & \textbf{554} & 1446 & nan \\
UVG-Jockey & 22.82 & \textbf{22.91} & 20.95 & nan & 0.147 & \textbf{0.124} & 0.201 & nan & \textbf{2349} & 4194 & 6426 & nan \\
UVG-ReadySteadyGo & 20.70 & 22.84 & 20.11 & \textbf{33.01} & 0.112 & 0.155 & 0.316 & \textbf{0.032} & 2832 & 3382 & 6347 & \textbf{902} \\
UVG-ShakeNDry & 24.68 & 26.59 & 24.43 & \textbf{36.68} & 0.111 & 0.077 & 0.158 & 0.0257 & 1400 & 2126 & 2896 & \textbf{689} \\
AverageValue & 24.47 & 30.41 & 23.28 & \textbf{34.96} & 0.104 & 0.088 & 0.184 & \textbf{0.022} & 2087 & 2343 & 4010 & \textbf{603} \\
\hline
\end{tabular}
}

     \caption{On the UVG dataset, a comparison of PSNR, LPIPS, and FVD among actual videos and different methods is conducted. NaN values indicate that the method cannot achieve the specified bpp value on this particular video.}\label{table2}
\end{table*}

\begin{figure*}[t]
\begin{center}
\begin{subfigure}[t]{0.64\columnwidth}
         \centering
         \includegraphics[width=\columnwidth]{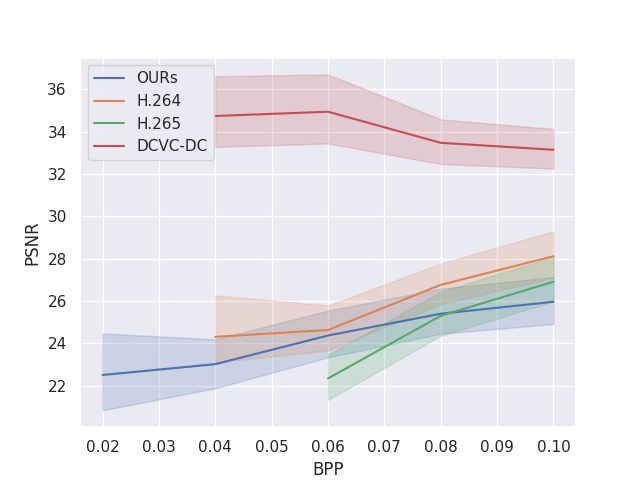}
         \caption{PSNR ($\uparrow$)}
         \label{fig:PSNR}
     \end{subfigure}
     \begin{subfigure}[t]{0.64\columnwidth}
         \centering
         \includegraphics[width=\columnwidth]{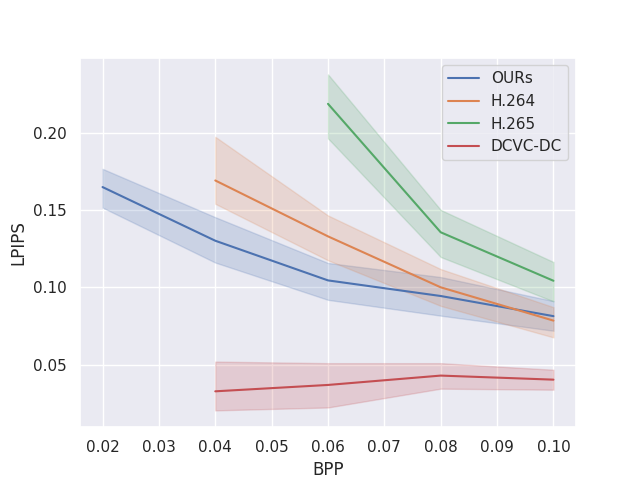}
         \caption{LPIPS ($\downarrow$)}
         \label{fig:LPIPS}
    \end{subfigure}
     \begin{subfigure}[t]{0.64\columnwidth}
         \centering
         \includegraphics[width=\columnwidth]{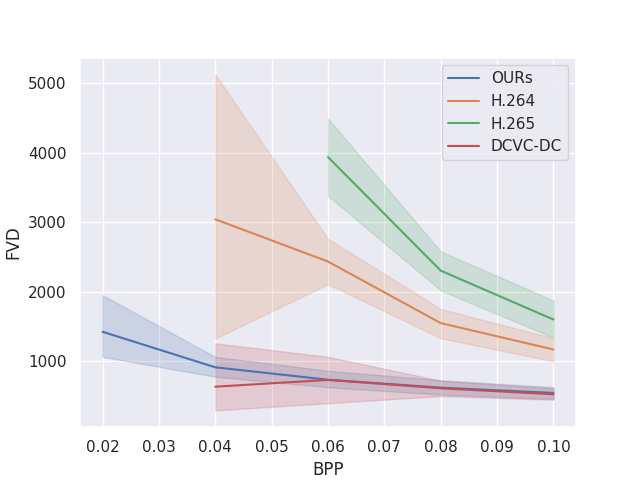}
         \caption{FVD ($\downarrow$)}
         \label{fig:FVD}
     \end{subfigure}
\caption{Rate-distortion (perception) performance on the Cityscapes dataset.}
\label{fig:RD}
\end{center}
\end{figure*}

\textbf{Implementation Details}: We employed pre-trained MCVD models from the training set and evaluated the model's performance on the test set.
For intra-frame compression, we employed the ELIC \cite{he2022elic}. We set the threshold for frame generation based on the LPIPS metric, with threshold values within the empirical range of [0.02, 0.30]. After processing each video sequence with the proposed compression scheme, we selected the optimal combination of ELIC compression quality and LPIPS threshold to represent the optimal performance of the proposed method.

\textbf{Baseline}: In addition to traditional video compression standards H.264 and H.265, we also compared the video compression performance with a SOTA neural compression method DCVC-DC\cite{li2023neural}. For each video sequence in the Cityscapes dataset, we transform the processed video frames into YUV420p format, then conduct compression and decompression operations.
In the case of grayscale datasets, due to the inherent support limitations of the libx264 codec, we convert the grayscale video frames into YUV420 format as YUV raw video for compression and decompression. For the evaluation, we only compute the metrics on the Y-component, therefore this approach could lead to higher compression video bit rates. 
For the H.264 and H.265 encoders, we control the generated video quality through the \textsc{CRF} parameter, which ranges from 0 to 51. 


\subsection{Results}
\begin{figure*}[ht]
\centering 
\includegraphics[scale=0.85]{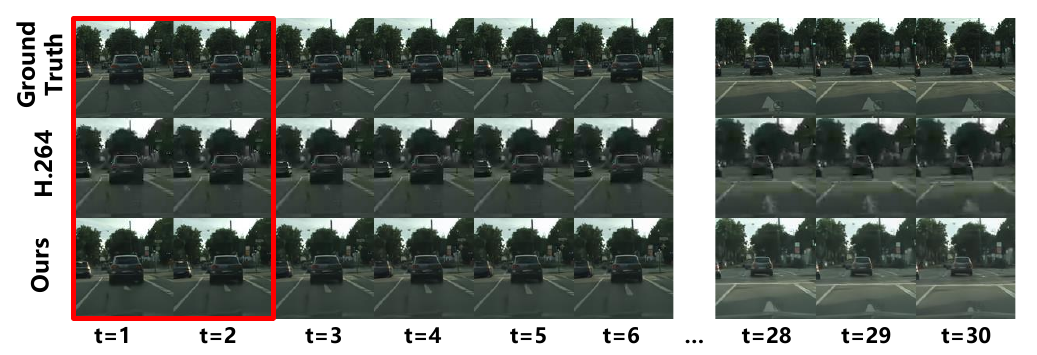}
\caption{\textbf{Visual comparison with state-of-the-art codec on Cityscape dataset.} We present the first 6 frames and last 3 frames between original video frames (top row), H.264 codec (second row) and the proposed compression scheme (bottom row). Both H.264 compression and the videos compressed by our model are controlled to have a bpp of 0.07.Our model utilizes the first two frames to autoregressively generate some following frames based on an LPIPS threshold of 0.16.}
\label{visualResult}
\end{figure*}

\begin{figure*}[ht]
\centering 
\includegraphics[scale=0.8]{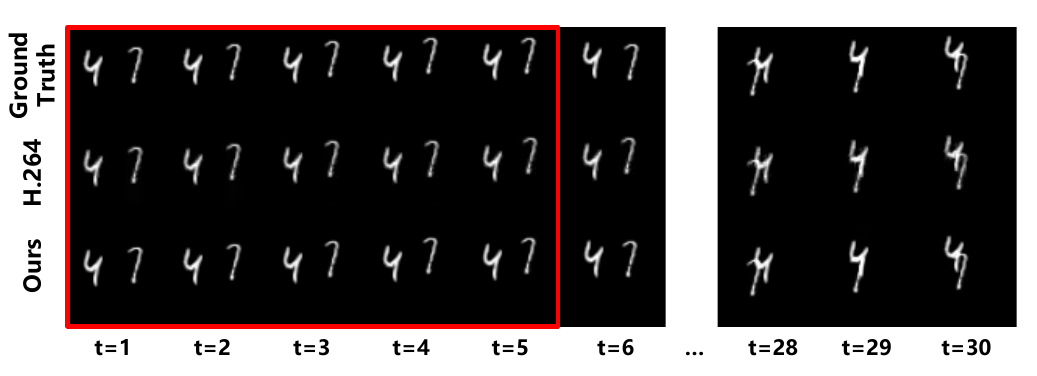}
\caption{\textbf{Visual comparison with state-of-the-art codec on SMMNIST dataset.} We present the first 6 frames and last 3 frames between original video frames (top row), H.264 codec (second row) and the proposed compression scheme (bottom row). Both H.264 compression and the videos compressed by our model are controlled to have a bpp  of 0.04,LPIPS threshold is 0.16. We generate frames conditioning on 5 frames. }
\label{smm_visualResult}
\end{figure*}

We visualize the results of our compression scheme in Figs.~\ref{visualResult} and \ref{smm_visualResult} of the Cityscapes and SMMNIST datasets, respectively. The first and second rows show the original frames and reconstructions, respectively, using H.264 as a baseline. The bottom row shows the reconstructed video frames using our method. It can be observed in Fig.~\ref{visualResult} that, our approach faithfully recovers the objects and textures while the standard codec suffers from blurriness, and in Fig.~\ref{smm_visualResult}, the last frames using our method still capture the shapes and positions of the original digits accurately, despite stochastic movements.

The rate-distortion-perception performances on the Citys-capes dataset are presented in Fig~\ref{fig:RD}. We report both the distortion metric PSNR (Fig.~\ref{fig:PSNR}) and the perceptual metrics LPIPS (Fig.~\ref{fig:LPIPS}) and FVD (Fig.~\ref{fig:FVD}) that are more aligned with human perception under different compression rates in the low bpp regime. 
The orange and green curves correspond to the H.264 and H.265 video compression standards, the red curve represents the experimental results of DCVC-DC, and the blue curve corresponds to the results of our method. In our experiments,we observed that while DCVC-DC performs well on high-resolution datasets, its compression capability is somewhat limited on low-resolution images, with a maximum achievable bitrate around 0.06. Both H.264 and H.265 can only achieve a bpp as low as 0.04 and 0.06 respectively, whereas our method achieves lower bpp (0.02) and consistently performs better in terms of the FVD metric. For the other perceptual measure LPIPS, the proposed method still outperforms H.265 and obtains similar performance to H.264. For the distortion metric PSNR, our method still shows comparable results concerning the standard codecs. 
In Table~\ref{table}, we report the distortion and perceptual qualities of the reconstructed video frames at the limiting bpp level (0.06) for the standard codecs. At the same bit rate, our method achieves better perceptual quality compared other methods.

In terms of the model's generalization ability, we also evaluated the model's performance on the UVG dataset in Table~\ref{table2}. Since the Cityscapes dataset has relatively monotonous colors and small motion variations, when faced with highly diverse samples, the generative capability, and hence, the compression performance, may degrade.

\subsection{Conclusion and Future Direction}
We proposed a novel predictive video compression scheme that achieves ultra-low bit rates with visually pleasing reconstruction results by combining a neural image compression model with a SOTA generative model.
In the proposed method, we sequentially decide a set of input frames to be encoded using neural codecs depending on whether the remaining frames can be generated by the decoder using previously encoded frames.
Each time, only a portion of the frames is encoded, and these frames are further used as conditional frames for the video generation model. Experimental results demonstrate that our model outperforms traditional video compression standards such as H.264 and H.265 in terms of perceptual qualities on datasets including SMMNIST and Cityscapes at ultra-low bpp. 
Our framework can serve as the basis for future works on predictive video compression employing different generative models at the receiver, and significantly reduce the required bitrate for video compression and transmission. 

We remark that, in the current implementation, the generation process has to be carried out both at the encoder and the decoder. 
The encoder employs the generation process to determine the quality of the generated video frames, which is then used to decide which video frames to be compressed as intra frames. 
This increases the encoding complexity of the proposed scheme; however, simpler prediction models will be explored as part of our future work to determine which frames should be compressed.

\bibliographystyle{IEEEbib}
\bibliography{refs}

\end{document}